\documentclass[12pt]{article}
\usepackage{amsmath}
\usepackage{cite}
\usepackage{graphicx}
\usepackage{color}
\begin{document}
\begin{flushright}
KANAZAWA-20-08\\
October, 2020
\end{flushright}
\vspace*{1cm}

\renewcommand\thefootnote{\fnsymbol{footnote}}
\begin{center} 
 {\Large\bf Low scale leptogenesis in a model with\\
 promising $CP$ structure}
\vspace*{1cm}

{\Large Daijiro Suematsu}\footnote[1]{e-mail:
~suematsu@hep.s.kanazawa-u.ac.jp}
\vspace*{0.5cm}\\

{\it Institute for Theoretical Physics, Kanazawa University, 
Kanazawa 920-1192, Japan}
\end{center}
\vspace*{1.5cm} 

\noindent
{\Large\bf Abstract}\\
We consider a simple extension of the standard model, which could give a solution to
its $CP$ issues through both the Peccei-Quinn mechanism and the Nelson-Barr 
mechanism.  Its low energy effective model coincides with the scotogenic model 
in the leptonic sector. Although leptogenesis is known not to work well at lower 
reheating temperature than $10^9$ GeV in simple seesaw  and scotogenic frameworks, 
such low reheating temperature could be consistent with both neutrino mass 
generation and thermal leptogenesis via newly introduced fields without referring 
to the resonance effect.  
An alternative dark matter candidate to axion is prepared as an indispensable ingredient 
of the model.

\newpage
\setcounter{footnote}{0}
\renewcommand\thefootnote{\alph{footnote}}

\section{Introduction}
Strong $CP$ problem \cite{strongcp} and origin of
$CP$ violating phases in the CKM matrix \cite{km} are  important issues unsolved 
in the standard model (SM). Recent neutrino oscillation experiments \cite{lepcp} 
suggest the existence of a $CP$ violating phase in the PMNS matrix also \cite{pmns}.
The formers are known to be solved by introducing Peccei-Quinn (PQ) symmetry 
$U(1)_{PQ}$ \cite{pq} and assuming complex Yukawa couplings for quarks \cite{km}.
Spontaneous breaking of the PQ symmetry brings about a pseudo 
Nambu-Goldstone boson called axion \cite{axion}. Since the PQ symmetry 
is explicitly broken to its discrete subgroup $Z_N$ via nonperturbative 
QCD effects, axion potential is lifted to cause $N$ degenerate vacua where 
the strong $CP$ problem is dynamically solved. Unfortunately, the domain wall 
appears to separate $N$ degenerate vacua \cite{pqwall}.
Since it is topologically stable for a case with $N\ge 2$,
its energy density overcloses the Universe.
On the other hand, the domain wall can decay in the model with $N=1$ \cite{stdw} 
and the problem is escapable. 
Such models can be realized for restricted field contents with special PQ charge 
assignment such as the KSVZ model \cite{ksvz}. 
Suitably constrained PQ charge could be also relevant to mass 
hierarchy of quarks and leptons \cite{dnone}.
Astrophysical and cosmological analyses require the PQ symmetry breaking 
scale $f_a$ to be within intermediate scales $10^{9-11}$ GeV \cite{fa}
as long as the PQ symmetry is broken after inflation.
It comes from the fact that axion mass and strength of its interaction 
with the SM fields are inversely proportional to $f_a$.\footnote{ 
The vacuum expectation value $v_{PQ}$ which breaks the PQ symmetry 
and the axion decay constant $f_a$ are related each other by $v_{PQ}=f_aN$
\cite{strongcp}. We use $f_a$ as its breaking scale since we consider only a 
case with $N=1$ throughout the paper.} 
An interesting feature of the model is that axion could be a dominant component 
of cold dark matter (DM) if $f_a$ takes a value near the 
upper bound mentioned above \cite{axiondm}. 
 
Inflation is considered to determine an initial condition of the hot 
Big-Bang Universe through the decay of inflaton.
If reheating temperature $T_R$ is high enough to satisfy $T_R>10^9$~GeV,
thermal leotogenesis based on out-of-equilibrium decay of a heavy right-handed 
neutrino \cite{leptg} is expected to work well in the seesaw framework 
for the neutrino mass \cite{di}.  
On the other hand, in the case $T_R<10^9$ GeV 
it cannot generate sufficient baryon number asymmetry through the 
thermal leptogenesis if we do not refer to resonance effects.
As long as we consider such a low reheating temperature, 
we have to consider some extension of the model.

In this paper, we propose an extended model which gives a solution to the $CP$ 
issues in the SM.
The strong $CP$ problem is solved by the PQ-mechanism 
and the $CP$ violating phases in the CKM  and PMNS matrices are 
spontaneously generated through Nelson-Barr mechanism for the strong 
$CP$ problem. Extra fields introduced for it play a crucial role 
to solve the difficulty in the low scale leptogenesis.
We show that the model can generate sufficient baryon number asymmetry through 
thermal leptogenesis in a consistent way with neutrino oscillation data for lower 
reheating temperature than $10^9$ GeV.
Although axion might not be a dominant component of DM, the model has another
DM candidate as an indispensable ingredient for the neutrino mass generation.    
The model is rather simple but we can treat several problems in the SM.

Remaining parts of the paper are organized as follows.
In section 2, we describe the model studied in this paper and discuss 
nature of the $CP$ symmetry in the model. 
In section 3, we first overview inflation and reheating considered in the model. 
After that, leptogenesis is discussed focusing on how sufficient lepton number 
asymmetry is generated at low reheating temperature less than $10^9$ GeV.  
We summarize the paper in section 4.
  
\section{A model with spontaneous $CP$ violation}
We consider an extension of the SM with a global $U(1)\times Z_4$ symmetry and 
several additional fields. 
The SM contents are assumed to have no charge of this global symmetry.
Added fermions are a pair of vector-like
down-type quarks $(D_L, D_R)$, a pair of vector-like charged leptons $(E_L, E_R)$,
and three right-handed singlet fermions $N_k~(k=1,2,3)$. Their representation and charge 
under  $[SU(3)_C\times SU(2)_L\times U(1)_Y]\times U(1)\times Z_4$ are 
given as\footnote{Similar models with vector-like extra fermions have 
been considered under different symmetry structure in several contexts \cite{mksvz,ps}. 
 } 
\begin{eqnarray}
&&D_L({\bf 3},~{\bf 1}, -\frac{1}{3})_{0,2}, \quad
D_R({\bf 3},~{\bf 1}, -\frac{1}{3})_{2,0}, \quad
E_L({\bf 1},~{\bf 1}, -1)_{0,2}, \quad
E_R({\bf 1},~{\bf 1}, -1)_{2,0}, \nonumber\\
&&N_k({\bf 1},~{\bf 1}, 0)_{1,1}.
\end{eqnarray}
We also introduce an additional doublet scalar $\eta$ and two singlet scalars
 $\sigma$ and $S$, whose representation and charge under the above symmetry 
are given as
\begin{equation}
\eta({\bf 1},~{\bf 2}, -\frac{1}{2})_{-1,1}, \quad
\sigma({\bf 1},~{\bf 1}, 0)_{-2,2}, \quad
S({\bf 1},~{\bf 1}, 0)_{0,2}.
\end{equation}
We note that this global $U(1)$ has color anomaly as one in the KSVZ 
model \cite{ksvz} and  it can play a role of the PQ symmetry.
Its charge assignment guarantees the domain wall number is one ($N=1$) 
so that the model can escape the domain wall problem.
We assume that the model is $CP$ invariant and then parameters contained
in Lagrangian are all real. 

The model is characterized by new Yukawa terms and
scalar potential which are invariant under the imposed symmetry
\begin{eqnarray}
-{\cal L}_Y&=&y_D\sigma\bar D_LD_R +y_E\sigma\bar E_LE_R+
\sum_{k=1}^3\Big[\frac{y_{N_k}}{2}\sigma \bar N_k^cN_k+ y_{d_k}S\bar D_L d_{R_k}
+ \tilde y_{d_k}S^\dagger\bar D_L d_{R_k} \nonumber\\ 
&+&y_{e_k}S\bar E_L e_{R_k} +\tilde y_{e_k}S^\dagger\bar E_L e_{R_k} +
\sum_{\alpha=1}^3 h_{\alpha k}^\ast\eta \bar\ell_\alpha N_k\Big]+ {\rm h.c.}, \nonumber \\
V&=&\lambda_1(\phi^\dagger\phi)^2+\lambda_2(\eta^\dagger\eta)^2
+\lambda_3(\phi^\dagger\phi)(\eta^\dagger\eta)+
\lambda_4(\phi^\dagger\eta)(\eta^\dagger\phi)+
\frac{\lambda_5}{2M_\ast}[\sigma(\phi^\dagger\eta)^2+{\rm h.c.}] \nonumber \\
&+&\kappa_\sigma(\sigma^\dagger\sigma)^2+\kappa_S(S^\dagger S)^2
+(\kappa_{\phi\sigma}\phi^\dagger\phi
+\kappa_{\phi\sigma}\eta^\dagger\eta)(\sigma^\dagger\sigma)
+(\kappa_{\phi S}\phi^\dagger\phi+\kappa_{\eta S}\eta^\dagger\eta) (S^\dagger S)
\nonumber \\
&+&\kappa_{\sigma S}\sigma^\dagger\sigma S^\dagger S 
+ m_\phi^2\phi^\dagger\phi+ m_\eta^2\eta^\dagger\eta
+m_\sigma^2\sigma^\dagger\sigma+m_S^2S^\dagger S +V_b,
\label{model}
\end{eqnarray} 
where $\ell_\alpha$ is a doublet lepton and $\phi$ is an ordinary doublet Higgs scalar.
$d_{R_k}$ and $e_{R_k}$ are the SM 
down-type quarks and charged leptons, respectively.
In eq.~(\ref{model}), we list dominant terms up to dimension five and 
$M_\ast$ is a cut-off scale of the model.
$V_b$ contains terms which are invariant under the global symmetry but 
violate the $S$ number such as $S^4$ and $S^{\dagger 4}$.
If the singlet scalars $\sigma$ and $S$ get vacuum expectation values (VEVs)
much larger than
the weak scale
\begin{equation}
\langle\sigma\rangle=we^{i\chi}, \qquad \langle S\rangle=ue^{i\rho},
\label{svev}
\end{equation}
we could have a low energy effective mode with spontaneous $CP$ violation.
The global symmetry $U(1)\times Z_4$ is broken to 
its diagonal subgroup $Z_2$ by these VEVs. This $Z_2$ guarantees the 
stability of a DM candidate as discussed later. 

Although we find $\chi=0$ due to the global $U(1)$ symmetry \cite{ssbcp},
$V_b$ can cause a non-zero $CP$ phase $\rho$ at a potential minimum 
through $\frac{\partial V_b}{\partial\rho}=0$. 
In fact,  if we assume as an example,
\begin{equation}
V_b=\alpha(S^4+S^{\dagger 4})+\beta\sigma^\dagger\sigma(S^2+S^{\dagger 2}),
\label{vb}
\end{equation}
$\cos 2\rho=-\frac{\beta}{4\alpha}\frac{w^2}{u^2}$ is obtained and its stability is found 
to require $\cos^22\rho<\frac{3}{4}$.
Since the global $U(1)$ symmetry works as the PQ symmetry, 
the VEV $w$ should be assumed to satisfy 
\begin{equation}
10^9~{\rm GeV}~{^<_\sim}~ w ~{^<_\sim}~10^{11} {\rm GeV}.
\end{equation}
The axion in this model is characterized by a coupling with photon 
$g_{a\gamma\gamma}=1.51~\frac{10^{-10}}{GeV}\left(\frac{m_a}{eV}\right)$ \cite{lmn}.
If we redefine each radial component of singlet fields around the vacuum 
as $\sigma=w+\frac{1}{\sqrt 2}\tilde\sigma$ and $S=u+\frac{1}{\sqrt 2}\tilde S$, the mass of  
$\tilde\sigma$ and $\tilde S$ is found to be 
$m^2_{\tilde\sigma}=4\kappa_\sigma w^2$ and $m^2_{\tilde S}=4\kappa_S u^2$, respectively.
In the example given by eq.~(\ref{vb}), mass of an orthogonal component to $\tilde S$ 
is found to be $12\alpha u^2\left(1-\frac{4}{3}\cos^22\rho\right)$. Its detail depends 
on the assumed $V_b$.

Here we note that the effective model after the symmetry breaking can give 
an explanation for origin of  $CP$ phases in the CKM  and PMNS 
matrices and the generation of neutrino masses. 
The former is based on a mass matrix which has been discussed by BBP \cite{bbp}
as a simple realization of Nelson-Barr mechanism \cite{nb}
for the strong $CP$ problem.
The latter is based on the fact that the leptonic sector coincides with the scotogenic 
model \cite{ma}. In the next part, we discuss them in some detail. 

\subsection{$CP$ violating phases in CKM and PMNS matrices}
The Yukawa couplings of down-type quarks and charged leptons shown in eq.~(\ref{model}) 
derive mass terms such as 
\begin{equation}
(\bar f_{Li}, \bar F_L){\cal M}_f
\left(\begin{array}{c} f_{R_j} \\ F_R \\ \end{array} \right)+{\rm h.c.}, \qquad
{\cal M}_f=\left(\begin{array}{cc}
m_{f_{ij}} & 0 \\ {\cal F}_{f_j} & \mu_F \\
\end{array}\right), 
\label{dmass}
\end{equation}
where ${\cal M}_f$ is a $4\times 4$ matrix. Characters $f$ and $F$ represent
$f=d,e$ and $F=D,E$ for down-type quarks and charged leptons, respectively.
 Each component of ${\cal M}_f$ is expressed as
 $m_{f_{ij}}=y_{f_{ij}}\langle\phi\rangle$, 
${\cal F}_{f_j}=(y_{f_j} ue^{i\rho} + \tilde y_{f_j} 
ue^{-i\rho})$ and $\mu_F=y_F w$.  
This mass matrix is found to have the same form proposed by BBP \cite{bbp}.
Since the global $U(1)$ symmetry works as the PQ symmetry and all parameters in the model 
are assumed to be real, $\bar\theta=\theta_{\rm QCD}+{\rm arg}({\rm det}{\cal M}_f)=0$ is 
satisfied even if radiative effects are taken into account 
after the spontaneous breaking of the $CP$ symmetry.

We consider diagonalization of a matrix ${\cal M}_f{\cal M}_f^\dagger$ 
by a unitary matrix such as
\begin{equation}
\left(\begin{array}{cc} A_f & B_f \\ C_f& D_f \\\end{array}\right)
\left(\begin{array}{cc} m_fm_f^{\dagger} & m_f{\cal F}_f^\dagger \\ 
  {\cal F}_fm_f^\dagger & \mu_F^2 +{\cal F}_f{\cal F}_f^\dagger \\
\end{array}\right)
\left(\begin{array}{cc} A_f^\dagger & C_f^\dagger \\ B_f^\dagger 
& D_f^\dagger \\\end{array}\right)=
\left(\begin{array}{cc} \tilde m^2_f & 0 \\ 0 & \tilde M^2_F \\\end{array}\right),
\label{mass}
\end{equation}
where a $3\times 3$ matrix $\tilde m^2_f$ is diagonal in which generation 
indices are abbreviated. Eq.~(\ref{mass}) requires
\begin{eqnarray}
   && m_fm_f^\dagger=A_f^\dagger \tilde m^2_f A_f+ 
C_f^\dagger \tilde M^2_FC_f, \qquad
  {\cal F}_fm_f^\dagger=B_f^\dagger m^2A_f+ D_f^\dagger M^2C_f, \nonumber \\
   && \mu_F^2+{\cal F}_f{\cal F}_f^\dagger=
B_f^\dagger \tilde m^2_f B_f+ D_f^\dagger \tilde M^2_F D_f.
\end{eqnarray}  
If $\mu_F^2+{\cal F}_f{\cal F}_f^\dagger$ is much larger than each 
components of ${\cal F}_fm_f^\dagger$, which means
$u, w\gg \langle \tilde\phi\rangle$,
we find that $B_f, C_f$ and $D_f$ can be approximated as
\begin{equation}
  B_f\simeq -\frac{A_fm_f{\cal F}_f^\dagger}{\mu_F^2
+{\cal F}_f{\cal F}_f^\dagger},
 \qquad C_f\simeq\frac{{\cal F}_f m_f^\dagger}{\mu_F^2
+{\cal F}_f{\cal F}_f^\dagger},
   \qquad D_f\simeq 1.
\end{equation}
These guarantee the unitarity of the matrix $A$ approximately. 
In such a case, it is easy to find 
\begin{eqnarray}
&&A_f^{-1}\tilde m_f^2A_f\simeq m_fm_f^\dagger -\frac{1}{\mu_F\mu_F^\dagger
+{\cal F}_f{\cal F}_f^\dagger}
(m_f{\cal F}_f^\dagger)({\cal F}_fm_f^\dagger), \nonumber \\
&&\tilde M_F^2\simeq \mu_F^2+ F_fF_f^\dagger.
\label{ckm}
\end{eqnarray}
The right-hand side of the first equation is an effective mass matrix of the ordinary 
fermions, which is derived through the mixing with the extra heavy fermions. 
Since its second term can have complex phases in off-diagonal 
elements as long as $y_{f_i}\not=\tilde y_{f_i}$ is satisfied, 
the matrix $A_f$ could be complex. Moreover, if the VEVs satisfy the condition
\begin{equation}
\langle\phi\rangle \ll w < u,
\label{uw}
\end{equation}
$\mu_F^2~{^<_\sim}~{\cal F}_f{\cal F}_f^\dagger$ could be realized for suitable 
Yukawa couplings $y_{f_j}, \tilde y_{f_j}$ and $y_F$.
In that case, the complex phase of $A_f$ in eq.~(\ref{ckm}) could have a substantial magnitude
because the second term is comparable with the first one.
The CKM matrix is determined as $V_{CKM}={O_L}^TA_d$ where $O_L$ is an
orthogonal matrix used for the diagonalization of an up-type quraks mass matrix.
Thus, the $CP$ phase of $V_{CKM}$ is caused by the one of $A_d$.
The same argument is applied to the leptonic sector
and we have the PMNS matrix as $V_{PMNS}=A_e^\dagger U$ where 
$U$ is an orthogonal matrix used for the diagonalization of a neutrino mass matrix 
which is discussed in the next part.
The Dirac $CP$ phase in the CKM matrix and the PMNS matrix can be 
explained by the same origin. A concrete example of them can be found in
Appendix of \cite{ps}.

\subsection{ Neutrino mass and DM} 
The leptonic sector of the effective model is characterized by the $Z_2$ invariant terms
\begin{eqnarray}
-{\cal L}_{\rm scot}&=&\sum_{k=1}^3
\left[\sum_{\alpha=1}^3h_{\alpha k}^\ast\bar\ell_\alpha\eta N_k +\frac{M_{N_k}}{2}\bar N_k^cN_k 
+ {\rm h.c.}\right] \nonumber \\
&+&\tilde m_\phi^2\phi^\dagger\phi+\tilde m_\eta^2\eta^\dagger\eta+
\tilde\lambda_1(\phi^\dagger\phi)^2+ \tilde\lambda_2(\eta^\dagger\eta)^2
+\tilde\lambda_3(\phi^\dagger\phi)(\eta^\dagger\eta)
+\lambda_4(\phi^\dagger\eta)(\eta^\dagger\phi) \nonumber \\
&+&\frac{\tilde\lambda_5}{2}\left[(\phi^\dagger\eta)^2+{\rm h.c.}\right].
\label{model0}
\end{eqnarray}
It is just the scotogenic model \cite{ma}.
After the spontaneous breaking due to the VEVs of $\sigma$ and $S$,
parameters in eq.~(\ref{model0}) are determined through integrating out 
$\tilde\sigma$ and $\tilde S$ by using the ones in 
eq.~(\ref{model}) as
\begin{eqnarray}
&&\tilde\lambda_1=\lambda_1-\frac{\kappa_{\phi\sigma}^2}{4\kappa_\sigma}
-\frac{\kappa_{\phi S}^2}{4\kappa_S}+
\frac{\kappa_{\sigma S}\kappa_{\phi\sigma}
\kappa_{\phi S}}{4\kappa_\sigma\kappa_S} \nonumber\\
&&\tilde\lambda_2=\lambda_2-\frac{\kappa_{\eta\sigma}^2}{4\kappa_\sigma}
-\frac{\kappa_{\eta S}^2}{4\kappa_S}+
\frac{\kappa_{\sigma S}\kappa_{\eta\sigma}
\kappa_{\eta S}}{4\kappa_\sigma\kappa_S} \nonumber\\
&&\tilde\lambda_3=\lambda_3-\frac{\kappa_{\phi\sigma}\kappa_{\eta\sigma}}
{2\kappa_\sigma}
-\frac{\kappa_{\phi S}\kappa_{\eta S}}{2\kappa_S}+
\frac{\kappa_{\sigma S}\kappa_{\phi\sigma}\kappa_{\eta S}
+\kappa_{\sigma S}\kappa_{\eta\sigma}\kappa_{\phi S}}{4\kappa_\sigma\kappa_S}\nonumber\\
&&\tilde\lambda_5=\lambda_5\frac{w}{M_\ast}, \qquad M_{N_k}=y_kw,
\end{eqnarray}
and
\begin{eqnarray} 
&&\tilde m_\phi^2=m_\phi^2+\left(\kappa_{\phi\sigma}+
\frac{\kappa_{\phi S}\kappa_{\sigma S}}{2\kappa_S}\right)w^2
+\left(\kappa_{\phi S}+
\frac{\kappa_{\phi\sigma}\kappa_{\sigma S}}{2\kappa_\sigma}\right)u^2, \nonumber\\
&&\tilde m_\eta^2=m_\eta^2+\left(\kappa_{\eta\sigma}+
\frac{\kappa_{\eta S}\kappa_{\sigma S}}{2\kappa_S}\right)w^2
+\left(\kappa_{\eta S}+
\frac{\kappa_{\eta\sigma}\kappa_{\sigma S}}{2\kappa_\sigma}\right)u^2.
\label{cpara}
\end{eqnarray}
Since $\eta$ is supposed to have no VEV, $Z_2$ is 
kept as an exact symmetry of the model. In this model, we assume
both $|\tilde m_\phi|$ and $\tilde m_\eta$ have values of $O(1)$~TeV although parameter 
tunings are required.
 
Neutrino mass is forbidden at tree level due to this $Z_2$ symmetry 
but it could be generated through one-loop diagrams with $\eta$ and $N_k$ 
in internal lines. Its formula is given as
\begin{equation}
{\cal M}^\nu_{\alpha\beta}\simeq\sum_{k=1}^3h_{\alpha k}h_{\beta k}\lambda_5\Lambda_k,  \qquad
\Lambda_k=\frac{\langle\phi\rangle^2}
{8\pi^2}\frac{1}{M_{N_k}}\ln\frac{M_{N_k}^2}{M_\eta^2},
\label{nmass}
\end{equation}
since $M_{N_k}\gg M_\eta$ is satisfied where 
$M_\eta^2=\tilde m_\eta^2+(\lambda_3+\lambda_4)\langle\phi\rangle^2$.
In order to make the point quantitatively clear, we assume a simple flavor 
structure for neutrino Yukawa couplings \cite{tribi}
\begin{equation}
h_{e i}=0, ~h_{\mu i}=h_{\tau i}\equiv h_i~~(i=1,~2); \quad  
h_{e 3}=h_{\mu 3}=-h_{\tau 3}\equiv h_3. 
\end{equation}
This realizes the tri-bimaximal mixing which gives a simple and good 0-th order 
approximation for the analysis of neutrino oscillation data and leptogenesis. 
If we impose the mass eigenvalues obtained from eq.~(\ref{nmass})       
to satisfy the squared mass difference required by the neutrino oscillation data, we find    
\begin{equation}
(h_1^2\Lambda_1+h_2^2\Lambda_2)|\lambda_5|=\frac{1}{2}\sqrt{\Delta m_{32}^2}, \qquad 
h_3^2\Lambda_3|\lambda_5|=\frac{1}{3}\sqrt{\Delta m_{21}^2}.
\end{equation}
If we assume $M_{2,3}=O(10^{7})$~GeV and $M_\eta=1$~TeV, 
we have $\Lambda_{2,3}=O(1)$~eV from the neutrino oscillation data \cite{pdg}, and 
it can be satisfied by $h_{2,3}=O(10^{-3})$ for $|\lambda_5|=10^{-3}$. 
In that case, $h_1$ can take a very small value compared with $h_{2,3}$. 

The axion may be difficult to be a dominant component of DM for $f_a<10^{10}$~GeV 
although it depends on the contribution from the axion string decay \cite{axiondm}.
However, fortunately, the model has another DM candidate, 
the lightest neutral component of $\eta$ with $Z_2$ odd parity.
It is known to be a good DM candidate which does not cause any contradiction 
with known experimental data as long as its mass is in the TeV range 
where the coannihilation can be effective \cite{idm,highmass,ks}.
In fact, if the couplings $\lambda_3$ and $|\lambda_4|$ take suitable values
 much larger than $|\lambda_5|$, both the DM abundance and the direct detection 
bound can be satisfied. Inelastic scattering between the neutral 
components mediated by $Z^0$ is also constrained through direct search experiments,
Its experimental bound can be satisfied for $|\lambda_5|~{^<_\sim}~10^{-5}$ \cite{inelas,ks-l}.   
We should note that this constraint is closely related to the neutrino mass 
generation as seen above. 
In the next section we examine the thermal leptogenesis at low reheating
temperature $T_R<10^9$ GeV taking account of the constraints from the neutrino 
oscillation data. 
      
\section{Leptogenesis at low reheating temperature}
We should note that the singlet scalar $S$ could play a role of inflaton in addition to give 
the origin of $CP$ violation in both quark and lepton sectors.
Since the $Z_4$ symmetry constrains its coupling with the Ricci scalar,
action relevant to the present inflation scenario is given 
in the Jordan frame as
\begin{equation}
S_J = \int d^4x\sqrt{-g} \left[-\frac{1}{2}M_{\rm pl}^2R -
\frac{\xi_1}{2} S^\dagger S R -\frac{\xi_2}{4}(S^2+S^{\dagger 2})R
+\frac{1}{2}\partial^\mu S^\dagger \partial_\mu S - V_S \right],
\label{inflag}
\end{equation}
where $M_{\rm pl}$ is the reduced Planck mass and $V_S$ stands for 
a corresponding part of the potential for $S$ and $S^\dagger$ in eq.~(\ref{model}). 
If we assume $\xi_1=-\xi_2$ is satisfied, the coupling of $S$ with the Ricci scalar
reduces to $\frac{1}{2}\xi S_I^2$ where $S=\frac{1}{\sqrt 2}(S_R+iS_I)$ and 
$\xi=\xi_1-\xi_2$.
It is well-known that a scalar field which couples non-minimally with the Ricci scalar 
in this way can cause inflation of the Universe \cite{nonm-inf}
and the idea has been applied to the Higgs scalar in the SM \cite{higgsinf} and its singlet scalar 
extensions \cite{sinfl}.\footnote{Inflation in the 
scotogenic model extended with singlet scalars has been discussed from 
several motivations \cite{ext-s}.}

If $S$ is assumed to evolve along a constant $\rho$ which is determined 
as a potential minimum $\frac{\partial V_b}{\partial\rho}=0$,
the radial component $\tilde S$ can be identified with inflaton in this model.
We suppose that $\tilde S$ takes a large field value of $O(M_{\rm pl})$ 
as an inflaton and other scalars have much smaller values than it during the inflation.
In that case, $V_S$ can be expressed as $V_S=\frac{\kappa_S}{4}\tilde S^4$.
After conformal transformation for a metric tensor in the Jordan frame, 
the corresponding potential $U$ and the canonically normalised inflaton $\chi$ 
in the Einstein frame is found to be written as 
\begin{equation}
U=\frac{\frac{1}{4}\kappa_S\tilde S^4}
{(1+\frac{\xi\sin^2\rho}{M_{\rm pl}^2}\tilde S^2)^2}, \qquad
\frac{d\chi}{d\tilde S}=\sqrt{6+\frac{1}{\xi\sin^2\rho}}\frac{M_{\rm pl}}{\tilde S}.
\end{equation} 
where we use $S_I=\tilde S\sin\rho$. 
If $\tilde S\gg \frac{M_{\rm pl}}{\sqrt\xi\sin\rho}$ is satisfied, $U$ takes a 
constant value $\frac{\kappa_S}{4\xi^2\sin^4\rho}M_{\rm pl}^4$
to cause inflation.
Since $\rho$ is supposed to give the origin of $CP$ 
violating phases in the CKM and PMNS matrices, $\sin\rho=O(1)$ 
seems to be favored from a view point of low energy effective theory.  
If we use the e-foldings number $N$,
slow-roll parameters in this inflation scenario is found to be expressed as 
$\epsilon=\frac{3}{4N^2}$ and $\eta=-\frac{1}{N}$ 
when $6\xi\sin^2\rho\gg 1$ is satisfied.
If we suppose such a case, the scalar spectral index $n_s$ and 
the tensor-to-scalar ratio $r$ are given as $n_s\sim 0.965$ 
and $r\sim 3.3\times 10^{-3}$ for $N=60$, 
which coincide well with the ones suggested by the Planck data \cite{planck18}.

The spectrum of density perturbation predicted by the inflation 
is known to be expressed as
\begin{equation}
{\cal P}(k)=A_s\left(\frac{k}{k_\ast}\right)^{n_s-1},  \qquad
A_s=\frac{U}{24\pi^2M_{\rm pl}^4\epsilon}\Big|_{k_\ast}. 
\label{power}
\end{equation}
If we use $A_s=(2.101^{+0.031}_{-0.034})\times 10^{-9}$ 
at $k_\ast=0.05~{\rm Mpc}^{-1}$ \cite{planck18}, we find the
Hubble parameter during the inflation to be $H_I=1.4\times 10^{13}
\left(\frac{60}{N}\right)$~GeV and the relation 
\begin{equation}
\kappa_S\simeq 1.49\times 10^{-6}\xi^2\sin^4\rho N^{-2},
\label{kap}
\end{equation}
which should be satisfied at the horizon exit time of the scale $k_\ast$.
Since the quartic coupling $\kappa_S$ is a free parameter in this model,
it allows $\xi$ to take a much smaller value 
in comparison with the one of the usual Higgs inflation.
For example, $\xi\sin^2\rho=O(10^2)$ realizes the observed value of 
$A_s$ for $N=60$ if $\kappa_S=O(10^{-6})$ is assumed.

After the end of inflation, the inflaton starts the oscillation around 
the vacuum $\langle S\rangle$.
During this oscillation, inflaton is expected to decay to light fields. 
If we assume the mass pattern 
\begin{equation}
2\tilde m_\eta<m_{\tilde S}< \tilde M_D, \tilde M_E ,
\end{equation}
the inflaton decay is expected to occur mainly through 
$\tilde S\rightarrow \eta^\dagger\eta$ at tree level.
The decay width could be estimated as
\begin{eqnarray}
\Gamma&\simeq& 
\frac{\kappa_{\eta S}^2}{16\pi \kappa_S}m_{\tilde S}
\sqrt{1-\frac{4\tilde m_\eta^2}{m_{\tilde S}^2}}.
\label{width}
\end{eqnarray}
After the inflaton decays to $\eta^\dagger\eta$,
the SM contents are thermalized through gauge interactions with $\eta$ 
immediately. The reheating temperature can be estimated as
\begin{equation}
T_R\simeq
1.6\times10^8\left(\frac{10^{-6}}{\kappa_S}\right)^{-1/4}
\left(\frac{\kappa_{\eta S}}{10^{-6.5}}\right)
\left(\frac{u}{10^{10}~{\rm GeV}}\right)^{1/2}~{\rm GeV}.
\end{equation}
As long as $\tilde M_D, \tilde M_E< T_R$ is satisfied, 
these fermions are also expected to be thermalized immediately 
through gauge interactions.
On the other hand, the right-handed neutrinos are considered to be 
thermalized  through the neutrino Yukawa couplings.
Here, we note that neutrino mass eigenvalues 
obtained from eq.~(\ref{nmass}) require $h_{2,3}=O(10^{-3})$ to
explain the neutrino oscillation data if $|\lambda_5|=O(10^{-3})$ and 
$M_{N_{2,3}}=O(10^7)$ GeV are assumed.
Since the decay width $\Gamma_{N_{2,3}}$ of $N_{2,3}$ and $T_R$ satisfy 
$\Gamma_{N_{2,3}}>H(T_R)$
and $T_R>M_{N_{2,3}}$ in such a case, $N_{2,3}$ are also expected to be 
thermalized through the inverse decay simultaneously at the reheating period.  
However, it is not the case for $N_1$ if its Yukawa coupling $h_1$ is 
much smaller than them. 

Baryon number asymmetry in the Universe \cite{baryon} is 
expected to be generated through leptogenesis \cite{leptg} in this model.
However, if reheating temperature is lower than $10^9$~GeV which corresponds to 
the lower bound of the PQ symmetry breaking,
leptogenesis might not work well since it coincides with the usually considered 
lower bound for successful leptogenesis in the seesaw framework \cite{di}. 
Since both the production of the right-handed neutrinos and the generation of
lepton number asymmetry through its out-of-equilibrium decay have to be 
caused only by neutrino Yukawa couplings there, 
it is difficult to yield a required lepton number asymmetry in a consistent 
way with the small neutrino mass generation. 
This situation does not change in the original scotogenic model either \cite{ks}.   

In the present model, the interaction between the right-handed 
neutrino $N_1$ and extra vector-like fermions mediated by $\tilde\sigma$
could change the situation in the similar way to the one discussed in \cite{lowlept}. 
In fact, scattering $\bar D_LD_R, \bar E_LE_R\rightarrow N_1N_1$ 
mediated by $\tilde\sigma$ could effectively produce the lightest 
right-handed neutrino $N_1$ in the thermal bath since $D_{L,R}$ and $E_{L,R}$ 
are in the thermal equilibrium as mentioned above. 
We can examine this possibility briefly.
We note first that the coupling constant $y_{N_1}$ should take a value 
of $O(10^{-2})$ at least to satisfy  $M_{N_1}<T_R$ 
since the VEV $w$ should be larger than  $10^9$ GeV.
Since the scattering becomes effective at the temperature 
$T$ such that $H(T)\simeq \Gamma_{F}$ where $\Gamma_{F}$ is the reaction rate of 
this scattering, $T>M_{N_1}$ should be satisfied to escape the Boltzmann suppression. 
Rough estimation of this condition gives 
\begin{equation}
T\simeq 6\times 10^8\left(\frac{y_F}{10^{-1.2}}\right)^2
\left(\frac{y_{N_1}}{10^{-2}}\right)^2 {\rm GeV},
\end{equation}
and we find that the present scenario could work for suitable $y_F$ 
since $T>M_{N_1}$ could be satisfied. 
We should note that this does not depend on the magnitude of 
the neutrino Yukawa coupling $h_1$ of $N_1$. 
It allows us to have successful leptogenesis even under $T_R<10^9$ GeV. 
Although this scattering process is expected to expand the allowed 
parameter space in the case $T_R>10^9$ GeV, we focus our present study
only on the low scale leptogenesis at $T_R<10^9$ GeV.

If $N_1$ is produced in the thermal bath successfully through the above mentioned
extra fermions scattering mediated by $\tilde\sigma$,
it decays to $\ell_\alpha\eta^\dagger$ in out-of-equilibrium 
through a strongly suppressed Yukawa coupling $h_1$. 
Since the decay is delayed until a period where the washout process caused 
by the inverse decay could be freezed-out and the generated lepton number 
asymmetry can be effectively converted to baryon number asymmetry 
through sphaleron processes.
We can check this scenario by solving Boltzmann equations for $Y_{N_1}$ and 
$Y_L(\equiv Y_\ell-Y_{\bar\ell})$, which are defined by using $\psi$ number 
density $n_\psi$ and entropy density $s$ as $Y_\psi=\frac{n_\psi}{s}$.
An equilibrium value of $Y_\psi$ is represented by $Y_\psi^{\rm eq}$.
The Boltzmann equations analyzed here are given as  
\begin{eqnarray}
&&\frac{dY_{N_1}}{dz}=-\frac{z}{sH(M_{N_1})}\left(\frac{Y_{N_1}}{Y^{\rm eq}_{N_1}}
-1\right)\left[ \gamma_D^{N_1}+ 
\left(\frac{Y_{N_1}}{Y^{\rm eq}_{N_1}}+1\right)\sum_{F=D,E}
\gamma_{F}\right], \nonumber \\
&&\frac{dY_{L}}{dz}=-\frac{z}{sH(M_{N_1})}\left[\varepsilon
\left(\frac{Y_{N_1}}{Y^{\rm eq}_{N_1}}-1\right)
\gamma_D^{N_1} -\frac{2Y_L}{Y_\ell^{\rm eq}}
\sum_{k=1,2,3}\left(\frac{\gamma_D^{N_k}}{4} +\gamma_{N_k}\right)\right],
\label{beq}
\end{eqnarray}
where $z=\frac{M_{N_1}}{T}$ and $H(T)$ is the Hubble parameter at temperature $T$.
$CP$ asymmetry for the decay of $N_1$ is expressed as $\varepsilon$. 
$\gamma_D^{N_k}$ is a reaction density for the decay 
$N_k\rightarrow \ell_\alpha\eta^\dagger$, and $\gamma_{N_k}$ is reaction density 
for lepton number violating scattering mediated by $N_k$ \cite{ks}. 
$\gamma_{F}$ represents a reaction density for the scattering 
$\bar D_LD_R,\bar E_LE_R\rightarrow N_1N_1$. 
As an initial condition at $T_R$, we assume $Y_{N_1}=Y_L=0$, and also 
$(D_L,D_R)$ and $(E_L,E_R)$ are in the thermal equilibrium. 

\begin{figure}[t]
\begin{center}
\begin{tabular}{ccc|ccc}\hline
&$\kappa_S$ & $\gamma$ & $\xi\sin^2\rho$ & 
 $T_R${\small  (GeV)} & $Y_B$   \\ \hline
(A)&$10^{-6}$ &  $\sqrt 2$&
 $49.1$ &  $1.6\times 10^8$ & $3.3\times 10^{-10}$ \\
(B)&$10^{-6}$ &2 &
 $48.9$  & $1.6\times 10^8$ &$1.6\times 10^{-10}$\\
(C)&$10^{-6}$ & $\sqrt 6$ &
 $48.4$  &$1.6\times 10^8$ &$9.6\times 10^{-11}$\\
(D)&$10^{-7}$ & $\sqrt 2$ &
  $15.4$  &$2.8\times 10^8$   & $4.9\times 10^{-10}$  \\
(E)&$10^{-7}$  & $2$ &
$14.6$  & $2.8\times 10^8$  &$2.9\times 10^{-10}$ \\
 (F)&$10^{-7}$ & $\sqrt 6$&
$12.7$  & $2.8\times 10^8$  &$2.0\times 10^{-10}$\\
\hline
\end{tabular}
\end{center}
\vspace*{2mm}
\footnotesize{{\bf Table 1}~ Expected values of nonminimal coupling, 
reheating temperature and baryon number asymmetry generated through leptogenesis. 
$\kappa_S$ is a value at an intermediate scale $w$ and $\xi$ is estimated at an 
inflation period by using one-loop RGEs. The condition for the substantial generation 
of the $CP$ phases in the CKM and PMNS matrices is satisfied for the used 
$\gamma$.} 
\end{figure}

Now we fix the VEVs as $w=10^9$~GeV and $u=10^{10}$~GeV as a typical
example. In numerical study of eq.~(\ref{beq}), we fix the relevant parameters at the
intermediate scale $w$ to guarantee the generation of $CP$ phases in the 
CKM and PMNS matrices through the mechanism discussed in the previous part. 
We require $\tilde M_F^2=\mu_F^2+{\cal F}_f{\cal F}_f^\dagger> \mu_F^2$ for it.
If we define $\gamma$ as $\tilde M_F=\gamma\mu_F$, it should satisfy  
$\gamma>1$.  
For simplification, we consider a case $y_f=(0,0,y)$ and  
$\tilde y_f=(0,\tilde y,0)$, which brings about to a relation 
$y^2+\tilde y^2=(\gamma^2-1)\frac{w^2}{u^2}y_F^2$ among Yukawa couplings
of extra fermions.
If $y_F$, $\gamma$ and $\tilde y/y$ are fixed,  $y$ and $\tilde y$ can be determined 
through this relation. 
We take $\tilde y/y=0.5$ in this study. 
By using these parameters, we can check the condition 
$\tilde M_D,~\tilde M_E,~M_{N_k}<T_R$. The stability of 
the inflaton potential at the inflation period
can be also examined by using the RGEs with  
contributions from the extra vector-like fermions.  
The $\kappa_S$ at the inflation period also allows us to find a value of 
$\xi$ through eq.~(\ref{kap}). 

As a typical value of parameters which could satisfy these conditions, 
we adopt $y_D=y_E=10^{-1.2}$, $\kappa_\sigma=10^{-4.5}$ and 
$\kappa_{S\eta}=10^{-6.5}$ at the intermediate scale $w$. 
We also fix parameters relevant to neutrino mass generation as
\begin{eqnarray}
y_{N_1}= 10^{-2}, \quad  y_{N_2}=2\times 10^{-2},  
\quad  y_{N_3}=4\times 10^{-2}, \quad  h_1=6\times 10^{-7}, 
\quad |\lambda_5|=10^{-3},
\label{para}
\end{eqnarray}
and $M_\eta=1$~TeV.   
Since the mass of $N_{k}$ is of $O(10^7)$ GeV for these parameters, 
neutrino Yukawa couplings $h_{2,3}$ are determined to be of  $O(10^{-3})$ 
by using the neutrino mass formula (\ref{nmass}) and the neutrino oscillation 
data \cite{pdg}.
If we assume a maximum $CP$ phase in the $CP$ asymmetry $\varepsilon$ in the
$N_1$ decay, $\varepsilon$ takes a value of $O(10^{-7})$ for this parameter setting.
Parameters $\kappa_S$ and $\gamma$ are fixed to the values given in
Table~1 at the intermediate scale $w$.

Solutions for the Boltzmann equations for different $\gamma$ values
are shown in Fig.~1, which confirms the present scenario to work.
The figures show that $Y_{N_1}$ reaches a value near its equilibrium one 
$Y_{N_1}^{\rm eq}$ through the scattering mediated by the extra fermions 
as expected. The out-of-equilibrium decay substantially occurs 
 at $z>10$ to generate the lepton number asymmetry.
This delay of the decay due to the small $h_1$ could make the lepton 
number asymmetry possible to escape the effective washout-out. 
Sufficient lepton number asymmetry is found to be produced before the 
sphaleron decoupling at $z_{EW}\sim\frac{M_{N_1}}{10^2~{\rm GeV}}$.
Since $\gamma$ becomes larger, the mass of extra fermions $\tilde M_F$ 
becomes larger to suppress the reaction density $\gamma_F$ 
due to the Boltzmann factor.
As a result, the $N_1$ number density generated through the scattering
becomes smaller and the resulting lepton number asymmetry also becomes 
smaller as shown in the figures.
If $h_1$ is much smaller, entropy produced through the decay of relic 
$N_1$ might dilute the generated lepton number asymmetry.
If $|\lambda_5|$ is taken to be smaller, Yukawa couplings $h_{2,3}$ become 
larger and the washout effect could remain effective until a later stage 
to reduce the resulting baryon number asymmetry.

\begin{figure}[t]
\begin{center}
\includegraphics[width=7.3cm,bb=6 50 366 302]{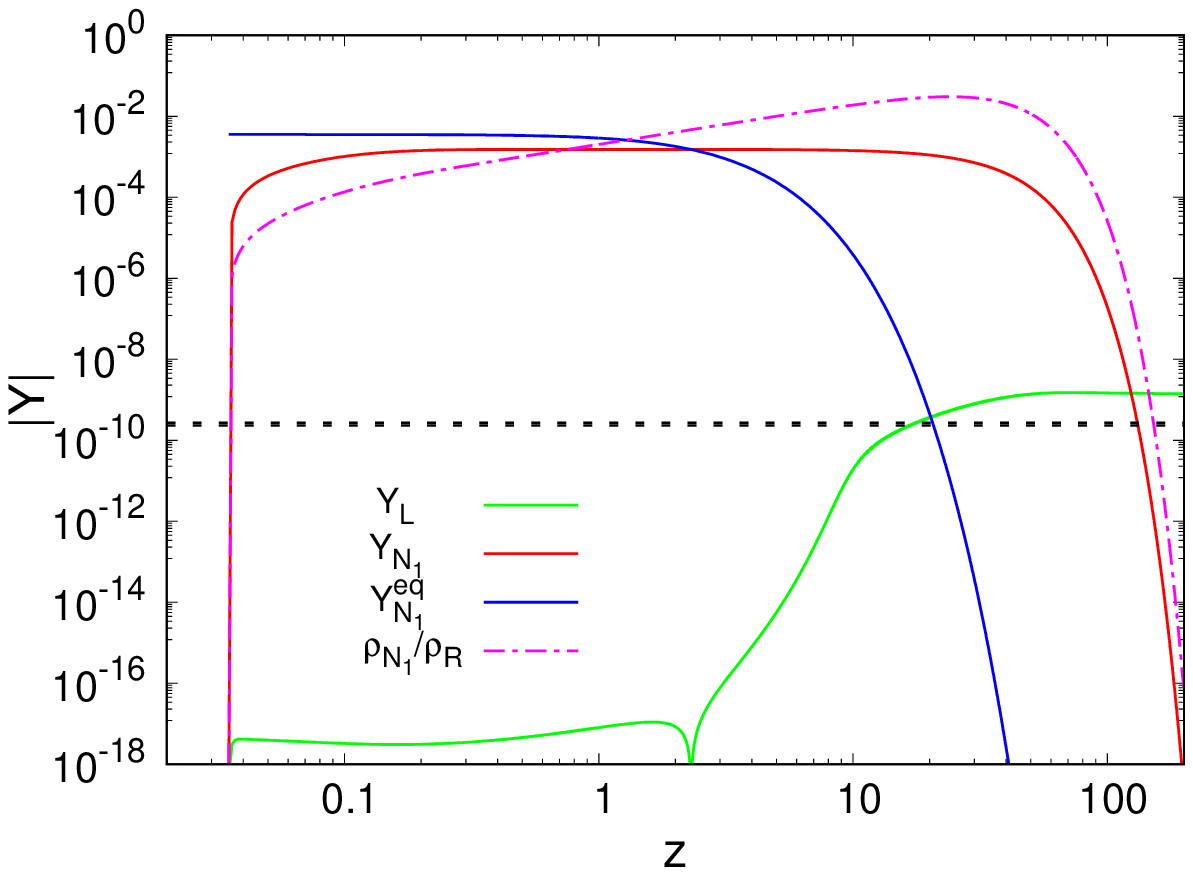}
\hspace*{3mm}
\includegraphics[width=7.3cm,bb=6 50 366 302]{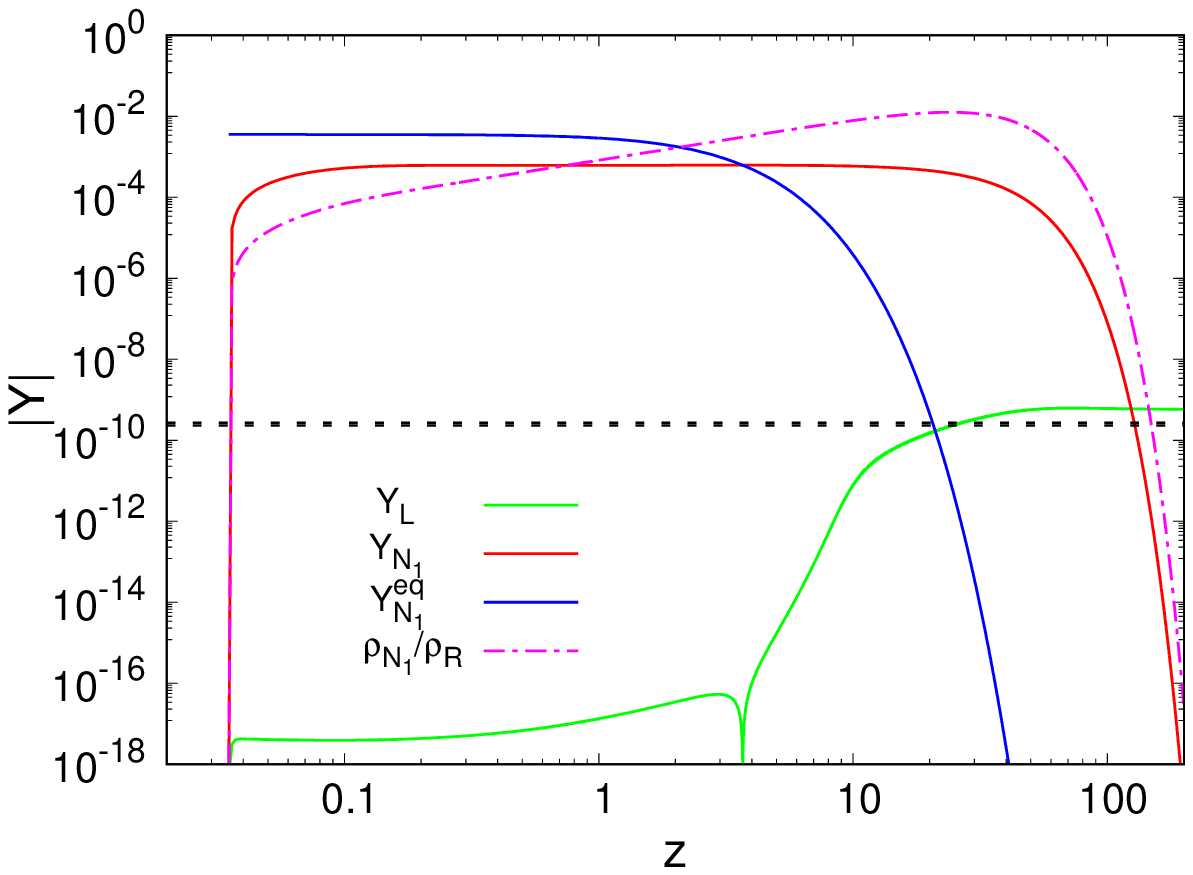}
\end{center}

\vspace*{7mm}
{\footnotesize {\bf Fig.~1}~Evolution of $Y_{N_1}$ and $Y_L$ obtained as a
solution of Boltzmann equations. The left panel  is one for the case (D) in Table 1
and the right panel is one for the case  (F). 
Initial values for them are fixed as $Y_{N_1}=Y_L=0$ at $z=z_R$.
$\rho_{N_1}/\rho_{\rm R}$ represents a ratio of the energy density of $N_1$
to the one of radiation.}
\end{figure}

In the last column of Table~1, the baryon number asymmetry generated 
for the assumed parameters is presented.
The difference of $Y_B$ for the same $\gamma$ can be also explained
as a result of suppression of the reaction density $\gamma_F$ by the Boltzmann
factor for extra fermions, which becomes smaller in the case with 
lower reheating temperature.
This result shows that the model with suitable parameters can generate a sufficient 
amount of baryon number asymmetry through leptogenesis 
although the reheating temperature is lower than $10^9$~GeV.
We should note that neutrino Yukawa couplings $h_{k}$ 
change their values depending on $w$ under the 
constraints of the neutrino oscillation data since the right-handed neutrino 
mass is generated through $M_{N_k}=y_kw$.
The extra fermion mass $\tilde M_F$ is also fixed by $w$ and $\gamma$.
Since $w$ is fixed as the lower bound of the $PQ$ symmetry breaking and 
$\gamma>1$ is imposed for the realization of the $CP$ phases in the CKM 
and PMNS matrices, we cannot expect that a scale of 
successful leptogenesis becomes much lower than the examples given here.
If we do not impose these conditions, a sufficient $Y_B$ could be
generated even for much lower reheating temperature. 

As mentioned already, it is difficult for the axion to be a dominant component 
of DM in the present parameter setting. However, the model has an alternative 
candidate, the lightest neutral component of $\eta$, as an indispensable 
component of the model. 
Thus, the model can give a simultaneous explanation for the strong $CP$ problem,
the origin of $CP$ phases in the CKM and PMNS matrices,  
leptogenesis and DM relic abundance for the reheating temperature lower 
than $10^9$ GeV.
 
\section{Summary}
We have proposed a model which can give an explanation for the strong $CP$
problem and the origin of the $CP$ phases in both the CKM and PMNS matrices.
It is a simple extension of the SM with vector-like extra fermions and several 
scalars. 
If the $CP$ is spontaneously broken in a singlet scalar sector at
an intermediate scale, it can be transformed to the CKM and PMNS matrices 
through mixings between the extra fermions and the ordinary quarks 
or the charged leptons. Their couplings are controlled by the imposed global symmetry.  
On the other hand, since the colored extra fermions play the same role 
as the ones in the KSVZ model for the strong $CP$ problem, 
the strong $CP$ problem can be solved through the PQ mechanism. 
After the symmetry breaking due to the singlet scalars, the leptonic sector of 
the model is reduced to the scotogenic model, which can explain the small 
neutrino mass and the DM abundance.
We have shown that the model has an interesting feature in addition to these.
The extra fermions could make the thermal leptogenesis possible to
generate the sufficient baryon number asymmetry even if the right-handed
neutrino mass is much lower than $10^9$~GeV, which is well-known lower bound 
of the right-handed neutrino mass for successful leptogenesis in the 
ordinary seesaw scenario.  
Although the axion cannot be a dominant component of DM in that case,
a neutral component of the inert doublet scalar  
can explain the DM abundance just as in the scotogenic model. 
It is remarkable that the model can explain various issues in the SM although
the model is rather simple.

\section*{Acknowledgements}
This work is partially supported by a Grant-in-Aid for Scientific Research (C) 
from Japan Society for Promotion of Science (Grant No. 18K03644).

\newpage
\bibliographystyle{unsrt}

\end{document}